\documentclass[aps, pre, reprint]{revtex4-2}
\usepackage{amsmath}
\usepackage{amssymb}
\usepackage{mathrsfs}
\usepackage{graphicx}
\usepackage{color}
\usepackage{hyperref}
\usepackage{xcolor}
\usepackage{soul}
\setstcolor{red}
\hypersetup{
    breaklinks = true,
    colorlinks = true,
    citecolor = {blue},
    urlcolor  = {blue},
    linkcolor = {blue}
}

\allowdisplaybreaks

\usepackage{xspace}
\usepackage{silence}
\newcommand{\eg}{e.g.,\xspace}
\newcommand{\ie}{i.e.,\xspace}
\newcommand{\pd}{\partial}
\newcommand{\ii}{\mathrm{i}}
\newcommand{\dd}{\mathrm{d}}
\newcommand{\ee}{\mathrm{e}}

\DeclareMathOperator{\re}{Re}
\DeclareMathOperator{\im}{Im}

\newcommand{\mc}[1]{\mathcal{#1}}
\newcommand{\msf}[1]{\mathsf{#1}}

\renewcommand{\vec}[1]{\mathbf{#1}}
\newcommand{\avrB}{\bar{\vec{B}}}
\newcommand{\avrn}{\bar{n}}

\newcommand{\eq}[1]{(\ref{#1})}
\newcommand{\Eq}[1]{Eq.~(\ref{#1})}
\newcommand{\Eqs}[1]{Eqs.~(\ref{#1})}
\newcommand{\Fig}[1]{Fig.~\ref{#1}}

\newcommand{\Refs}[1]{Refs.~\onlinecite{#1}}
\newcommand{\myRef}[1]{Ref.~\onlinecite{#1}}
\ifx\Ref\undefined
\newcommand{\Ref}[1]{\myRef{#1}}
\else
\renewcommand{\Ref}[1]{\myRef{#1}}
\fi

\newcommand{\note}[1]{{#1}}

\begin{document}
\title{Spin Hall effect of radiofrequency waves in magnetized plasmas}
\author{Yichen Fu}
\email{yichenf@princeton.edu}
\author{I.~Y. Dodin}
\author{Hong Qin}
\affiliation{
    Princeton Plasma Physics Laboratory, Princeton, NJ 08543\\
    Department of Astrophysical Sciences, Princeton University, Princeton, NJ 08544
}
\begin{abstract}
In inhomogeneous media, electromagnetic-wave rays deviate from the trajectories predicted by the leading-order geometrical optics. This effect, called the spin Hall effect of light, is typically neglected in ray-tracing codes used for modeling waves in plasmas. Here, we demonstrate that the spin Hall effect can be significant for radiofrequency waves in toroidal magnetized plasmas whose parameters are in the ballpark of those used in fusion experiments. For example, an electron-cyclotron wave beam can deviate by as large as ten wavelengths ($\sim 0.1\,\text{m}$) relative to the lowest-order ray trajectory in the poloidal direction. We calculate this displacement using gauge-invariant ray equations of extended geometrical optics, and we also compare our theoretical predictions with full-wave simulations.
\end{abstract}

\maketitle

\section{Introduction}

Precise modeling of radiofrequency (RF) waves in magnetic-fusion experiments is essential for many applications, including cyclotron heating, current drive \cite{ref:fisch87, prater2008benchmarking, figini2012benchmarking}, and suppression of tearing modes \cite{reiman1983suppression,petty2004complete}. Waves with short wavelengths, particularly those in the electron-cyclotron and lower-hybrid frequency range, are commonly modeled with ray-tracing codes \cite{kritz1982ray, pereverzev1998beam, smirnov2001genray, poli2001torbeam, farina2007quasi, marushchenko2014ray}, which are based on geometrical optics (GO) \cite{kravtsov1990geometrical, tracy2014ray}. In the traditional, lowest-order GO, the evolution of the ray coordinate $\mathbf{x}$ and momentum $\mathbf{k}$ (wavevector) are governed by the Hamiltonian that equals the local dispersion function and ignores the local gradients of the medium parameters \cite{stix1992waves}. However, the evolution of the wave polarization in an inhomogeneous medium produces corrections that make rays deviate from the trajectories predicted by this lowest-order Hamiltonian \cite{littlejohn1991geometric, weigert1993diagonalization}. 
This effect, which is similar to the spin-orbital interaction in quantum mechanics and the spin-orbital-like coupling in atomic systems \cite{zygelman1990non,zygelman2013non}, is known as the spin Hall effect (SHE) of light. It is being actively studied as an important photonic effect in various physical systems \cite{onoda2004hall, bliokh2006conservation, bliokh2008geometrodynamics, bliokh2015spin, ruiz2015lagrangian, ruiz2015first, ling2017recent, perez2021manifestation, yamamoto2018spin, oancea2020gravitational}. Plasma was recently identified as a medium where the SHE manifests \cite{ruiz2017extending, ruiz2017geometric, dodin2019quasioptical}.  But the importance of the SHE for practical plasma applications has not been explored, and the common ray-tracing codes used in fusion research ignore the SHE entirely \cite{foot:quasi}. In this paper, we show that the SHE can be significant for RF waves in magnetized plasmas whose parameters are in the ballpark of those used in fusion experiments. The results reported enrich the understanding of the SHE and demonstrate the importance of an interdisciplinary approach to wave effects found in various media.

We calculate the SHE using equations of ``extended GO'' (XGO) as formulated in \Refs{dodin2019quasioptical, ruiz2017extending}. By comparing XGO predictions with two-dimensional (2-D) full-wave (FW) simulations, we show that the ray equations that account for the SHE describe waves in fusion plasmas more accurately than the traditional ray equations. For three-dimensional (3-D) toroidal plasmas, we show that electromagnetic waves in the electron-cyclotron frequency range can deviate by as large as ten wavelengths in the poloidal plane relative to the GO predictions. Displacements of this size can be important in practice, for example, because processes like collisionless absorption and mode conversion can be very sensitive to the spatial distribution of the wave energy. 

Our paper is organized as follows. In Sec.~\ref{sec:basic_eq}, we introduce basic theory. In Sec.~\ref{sec:application}, we apply this theory to waves in cold magnetized plasma, specifically, in a plasma slab (Sec.~\ref{sec:slab}) and toroidal plasma (Sec.~\ref{sec:toroidal}). In Sec.~\ref{sec:conclusion}, we conclude our results. Auxiliary calculations are presented in appendices.

\section{Basic equations \label{sec:basic_eq}}

Let us start by briefly restating the derivation of the XGO ray equations \cite{ruiz2017geometric, dodin2019quasioptical}. Consider a multicomponent wave field $\Psi$ on a given space (or spacetime) $x^\mu$ and suppose that this wave is governed by an equation of the form $\hat{D}\Psi = 0$, where $\hat{D}$ is some linear dispersion operator generally of integro-differential form. Assume that the wave has an eikonal form $\Psi(x^{\mu}) = \ee^{\ii\theta(x^{\mu})}\psi(x^{\mu})$. Here, the scalar function $\theta$ is a fast real phase and $\bar{k}_{\mu} \doteq \partial_\mu \theta$ is the associated wavevector, which is generally a field on $x^\mu$. (The symbol $\doteq$ denotes definitions, $\pd_\mu \doteq \pd/\partial x^{\mu}$, and $\pd^\mu \doteq \pd/\pd k_{\mu}$.) The multicomponent function $\psi$ represents a slow complex envelope governed by $\hat{\mathcal{D}} \psi = 0$, where $\hat{\mathcal{D}} \doteq \ee^{-\ii\theta}\hat{D} \ee^{\ii\theta}$. We assume that the least scale $L$ of the envelope dynamics is much smaller than the local wavelength $\lambda \doteq 2\pi/\bar{k}$, yielding a small parameter $\epsilon \doteq \lambda/L \ll 1$ (the ``GO parameter''). Then, assuming Euclidean or pseudo-Euclidean (\eg Minkowski) coordinates for simplicity, one obtains \cite{dodin2019quasioptical}
\begin{gather}
    \label{eq:approx_dispersion_operator}
    \hat{\mathcal{D}} \approx \mathsf{D}(x^{\mu}, \bar{k}_{\mu}(x^\mu))
    - \ii \msf{V}^\mu \pd_\mu - \ii(\pd_\mu \msf{V}^\mu)/2.
\end{gather}
Here, the ``local dispersion tensor'' $\mathsf{D}$, which is a matrix function on the ray phase space $(x^\mu, k_\mu)$, is the Weyl symbol of $\hat{D}$, and $\msf{V}^\mu \doteq (\pd^\mu \mathsf{D}(x^{\mu}, k_{\mu}))|_{k_\mu = \bar{k}_\mu(x^\mu)}$. We assume that the Hermitian part of the dispersion tensor, $\mathsf{D}_{\rm H} \doteq (\mathsf{D} + \mathsf{D}^\dag)/2$, is $\mc{O}(1)$, while its anti-Hermitian part, $\mathsf{D}_{\rm A} \doteq (\mathsf{D} - \mathsf{D}^\dag)/(2\ii)$, is $\mc{O}(\epsilon)$, so one can use $\msf{V}^\mu(x^\mu) \approx (\pd^\mu \mathsf{D}_{\rm H}(x^{\mu}, k_{\mu}))|_{k_\mu = \bar{k}_\mu(x^\mu)}$. Then, $\mathsf{D}_{\rm A}$ determines local damping, while the wave propagation is determined entirely by $\mathsf{D}_{\rm H}$, namely, as follows. 

To the zeroth order in $\epsilon$, \Eq{eq:approx_dispersion_operator} gives
\begin{gather}\label{eq:zero}
\mathsf{D}_{\rm H}(x^{\mu}, \bar{k}_{\mu}(x^\mu))\psi(x^\mu) = 0,
\end{gather}
so $\psi$ is the eigenvector of $\mathsf{D}_{\rm H}$ corresponding to an eigenvalue $\Lambda$ that is zero on the solution at $\epsilon \to 0$. To eliminate mode conversion, which makes the analysis more complicated \cite{dodin2019quasioptical}, let us assume that $\mathsf{D}_{\rm H}(x^{\mu}, k_{\mu})$ has only one eigenvalue that is zero at $k_\mu = \bar{k}_\mu(x^\mu)$; we call it an active mode. Then, \Eq{eq:zero} can be restated as follows. Consider the corresponding unit eigenvector $\eta$,
\begin{gather}\label{eq:eta}
\mathsf{D}_{\rm H}\eta = \Lambda \eta, 
\quad
\eta^\dag \eta = 1,
\quad
\Lambda = \eta^\dag \mathsf{D}_{\rm H}\eta,
\end{gather}
where all quantities are considered as functions of $(x^{\mu}, k_{\mu})$. Then, $\Lambda(x^{\mu}, \bar{k}_{\mu}(x^\mu)) = 0$ serves as an approximate dispersion relation, and $\psi(x^\mu) = \eta(x^\mu, \bar{k}_\mu(x^\mu)) a(x^\mu)$, where $a(x^\mu)$ is a scalar amplitude. To separate the phase dynamics from the amplitude dynamics completely, we attribute the \textit{whole} phase that the envelope may have to $\theta$ and $\eta$, so the function $a$ is real by definition. We note that $\eta$ is defined only up to $\ee^{\ii \varphi}$, where $\varphi$ is real and slow but otherwise arbitrarily. This constitutes a U(1) gauge symmetry of XGO, with $\varphi$ being the gauge potential (see below).

To the first order in $\epsilon$, \Eq{eq:approx_dispersion_operator} gives $\psi(x^\mu) = \eta(x^\mu, \bar{k}_\mu(x^\mu)) a(x^\mu) + \psi_\perp(x^\mu)$, where $\psi_\perp = \mc{O}(\epsilon)$ is perpendicular to $\eta$. Then, projecting \Eq{eq:approx_dispersion_operator} on $\eta$ gives \cite{dodin2019quasioptical} 
\begin{gather}
(\Lambda - U)a - \ii[V^\mu \pd_\mu + (\pd_\mu V^\mu)/2 - \Gamma]a = 0.
\end{gather}
Here, $\Lambda$ is as in \Eq{eq:eta}, $\Gamma \doteq \eta^\dag \mathsf{D}_{\rm A}\eta$, $U \doteq \im[(\pd_\mu \eta^\dag) \msf{V}^\mu \eta]$, and $\eta = \eta(x^\mu, \bar{k}_\mu(x^\mu))$, so $\pd_\mu$ applies to both its arguments. Since $a$ is real by definition, the imaginary part of \Eq{eq:approx_dispersion_operator} gives an amplitude equation $[V^\mu \pd_\mu + (\pd_\mu V^\mu)/2 - \Gamma]a = 0$, and the real part of \Eq{eq:approx_dispersion_operator} gives a dispersion relation $\mc{H}(x^\mu, \bar{k}_\mu(x^\mu)) = 0$, where $\mc{H} \doteq \Lambda - U$ and~\cite{dodin2019quasioptical}
\begin{gather}
    U = U_0 - \mathsf{A}_{\mu}^{(x)}\,\pd^\mu\Lambda + \mathsf{A}_{(k)}^{\mu}\,\pd_\mu \Lambda,\\
    U_0 \doteq \im \left[ \left(\partial^\mu\eta^{\dag}\right) \mathsf{D}_{\mathrm{H}} \left(\partial_\mu\eta\right) \right],
    \label{eq:U0}\\
    \mathsf{A}_{\mu}^{(x)} \doteq \im \left( \eta^{\dag} \partial_\mu\eta \right),
    \quad
    \mathsf{A}_{(k)}^{\mu} \doteq \im\left(\eta^{\dag} \partial^\mu\eta \right).
    \label{eq:As}
\end{gather}
As usual \cite{tracy2014ray}, the corresponding ray equations are
\begin{gather}
\dot{x}^\mu = \pd^\mu \mc{H},
\quad
\dot{k}_\mu = - \pd_\mu \mc{H},
\end{gather}
where the dot denotes $\dd/\dd\tau$, $\tau$ is a parameter along the ray, and $k_\mu$ represents $\bar{k}_\mu$ (the bar is omitted from now on for brevity). These correspond to the canonical Lagrangian $\mc{L} = k_\mu \dot{x}^\mu - \mc{H}$. But note that $\dot{x}^\mu = \pd^\mu \Lambda + \mc{O}(\epsilon)$, $\dot{k}_\mu = -\pd_\mu \Lambda + \mc{O}(\epsilon)$, $ \mathsf{A}_{\mu}^{(x)} = \mc{O}(\epsilon)$, and $\mathsf{A}^{\mu}_{(k)} = \mc{O}(\epsilon)$, so $U \approx U_0 - \dot{x}^{\mu}\mathsf{A}_{\mu}^{(x)} - \dot{k}_{\mu}\mathsf{A}_{(k)}^{\mu} + \mc{O}(\epsilon^2)$. Omitting $\mc{O}(\epsilon^2)$, which is negligible within the assumed accuracy, one arrives at an equivalent noncanonical Lagrangian
\begin{gather}
\mc{L} = \left[\dfrac{k_\mu}{2} - \mathsf{A}_{\mu}^{(x)}\right]\dot{x}^\mu - \left[\dfrac{x^\mu}{2} + \mathsf{A}^{\mu}_{(k)}\right]\dot{k}_\mu - (\Lambda - U_0),
\label{eq:L}
\end{gather}
where an insignificant full derivative $\dd(k_\mu x^\mu/2)/\dd\tau$ has been omitted. The functions $\msf{A}$ define the Berry connections \cite{simon1983holonomy}. It plays the role of the Abelian vector potential in the present context. The corresponding Euler--Lagrange equations are as follows:
\begin{subequations}
\label{eq:rays}
\begin{align}
\dot{x}^{\mu} & =+ \partial^\mu(\Lambda-U_{0}) + {\mathsf{F}^{\mu}}_{\nu}\dot{x}^{\nu}+\mathsf{F}^{\mu\nu}\,\dot{k}_{\nu},
\label{eq:XGO1}\\[5pt]
\dot{k}_{\mu} & =- \partial_\mu(\Lambda-U_{0}) - {\mathsf{F}_{\mu}}^{\nu}\dot{k}_{\nu}-\mathsf{F}_{\mu\nu}\,\dot{x}^{\nu},
\label{eq:XGO2}
\end{align}
\end{subequations}
where the functions $\msf{F}$ (with various indices) are understood as the Berry curvatures:
\begin{subequations}
    \label{eq:Fs}
    \begin{align}
    {\mathsf{F}^{\mu}}_{\nu} & \doteq 2\im [ ( \partial^\mu\eta^{\dag}) \partial_\nu\eta ],\,\,
    \mathsf{F}^{\mu\nu}\doteq 2\im [ ( \partial^\mu\eta^{\dag})  \partial^\nu\eta ],
    \label{eq:berry_curv1}\\
    {\mathsf{F}_{\mu}}^{\nu} & \doteq 2\im[ (\partial_\mu\eta^{\dag}) \partial^\nu\eta ],\,\,
    \mathsf{F}_{\mu\nu}\doteq 2\im[(\partial_\mu\eta^{\dag}) \partial_\nu\eta].
    \label{eq:berry_curv2}
    \end{align}
\end{subequations}
This result generalizes the corresponding equations in \Refs{perez2021manifestation}  to  arbitrary dispersive operators $\hat{D}$. In particular,  one can show (see Appendix~\ref{sec:gauge}) that \Eqs{eq:rays} are gauge-invariant, \ie unaffected by a variable transformation $\eta \to \ee^{\ii\varphi}\eta$. Still, \Eqs{eq:U0} and \eq{eq:Fs} can be inconvenient for numerical integration. Assuming that $\eta$ is provided by a generic eigensolver, having its phase arbitrary makes $\eta(x^\mu, k_\mu)$ discontinuous, resulting in large $\msf{F}$. This undermines both the accuracy of the perturbation model and the numerical stability of the integrator. A more practical form of $U_0$ and $\msf{F}$ is derived in Appendix~\ref{sec:gauge_invariant_formula} and generalize the corresponding formulas from \Refs{berry1984quantal, perez2021manifestation}:
\begin{subequations}
\label{eq:UF}
\begin{gather}
U_{0} = \mathrm{Im} \sum_m\dfrac{\eta^{\dag}(\partial^\mu\msf{D}_{\rm H})\eta_{m}\eta_{m}^{\dag}(\partial_\mu\msf{D}_{\rm H})\eta}{\Lambda_m},\label{eq:gauge_invariant_U}
\\
\mathsf{F}^{\mu\nu} = 2\mathrm{Im} \sum_m\dfrac{\eta^{\dag}(\partial^\mu\msf{D}_{\rm H})\eta_{m}\eta_{m}^{\dag}(\partial^\nu\msf{D}_{\rm H})\eta}{\Lambda_m^{2}}.\label{eq:gauge_invariant_F}
\end{gather}
\end{subequations}
Lowering the indices $\mu$ and $\nu$ on the right-hand side of \Eq{eq:gauge_invariant_F} yields the corresponding components of $\msf{F}$ with lower and mixed indices. Here, $\eta_m$ and $\Lambda_m$ are the unit eigenvectors and the corresponding eigenvalues of $\msf{D}_{\rm H}$ that correspond to modes with $\Lambda_m \ne \Lambda$; we call them passive modes. These equations have the benefit of ``numerical gauge invariance'' in that changing $\eta \to \ee^{\ii\varphi}\eta$ leaves both $U_0$ nor $\mathsf{F}$ intact. It is also seen from \Eqs{eq:UF} that the SHE is amplified when one or more of the passive modes is in resonance with the active mode, namely, $\Lambda_m \approx 0$. In this regime, one can also expect mode conversion, \ie tunneling of the wave action between separate dispersion surfaces. However, this tunneling scales with $\Lambda_m$ exponentially \cite{tracy2014ray, my:xo}, while the SHE scales with $\Lambda_m$ algebraically, so it can be amplified at small $\Lambda_m$ while the mode conversion remains negligible. 

The SHE in the present study originates from the internal degree of freedom associated with vector components of the photon field in plasmas. It is also similar to the spin-orbital-like coupling that appears in the non-Abelian gauge field theory for electron wave functions in atomic systems \cite{zygelman1990non,zygelman2013non}, but the mechanism there is different. Another difference is that the gauge potential in the present context is Abelian, since we only study here the propagation of one single mode away from the mode-conversion region, and the eigenmode is non-degenerate. For the plasma waves near the mode conversion region, the effects of eigenmode degeneracy are important \cite{ruiz2017extending}. Then a non-Abelian gauge theory similar to that in atomic systems \cite{zygelman1990non,zygelman2013non} is necessary.

\section{Waves in cold magnetized plasma \label{sec:application}}

\subsection{Basic equations}

Let us apply the above results to a cold-plasma model, which is commonly used for ray tracing in fusion devices \cite{prater2008benchmarking}. For simplicity, let us neglect the ion response, plasma flows, and dissipation, and let us also assume that the waves have a fixed frequency $\omega$, as usual. Then, the linear-wave equations can be written as $\hat{D}\Psi = 0$, where $\mathsf{D} = \mathsf{D}_{\mathrm{H}} = H(\mathbf{x}, \mathbf{k}) - \omega$, 
\begin{gather}\label{eq:Hham}
\hat{H}(\mathbf{x}, -\ii\partial_{\mathbf{x}}) =
\begin{pmatrix}
-\ii\boldsymbol{\Omega}(\mathbf{x})\times & \ii\omega_{\mathrm{p}}(\mathbf{x}) & 0\\
-\ii\omega_{\mathrm{p}}(\mathbf{x}) & 0 & \ii c\,\partial_{\mathbf{x}}\times\\
0 & -\ii c\,\partial_{\mathbf{x}}\times & 0
\end{pmatrix},
\end{gather}
and $\Psi(t,\mathbf{x}) = (\mathbf{v}, \mathbf{E}, \mathbf{B})^\intercal$ is a 9-dimensional column vector (the symbol $^\intercal$ denotes transposition) that includes the properly normalized velocity $\vec{v}$, the wave electric field $\vec{E}$, and the wave magnetic field $\vec{B}$ \cite{my:qc1}. Also, $\times$ denotes vector product, as usual; $\omega_{\mathrm{p}}(\mathbf{x})=[4\pi q^{2}\avrn(\mathbf{x})/m]^{1/2}$ is the plasma frequency; $q$, $m$, and $\avrn(\mathbf{x})$ are the electron charge, mass, and background density, correspondingly; $\boldsymbol{\Omega}(\mathbf{x}) = q\avrB(\mathbf{x})/(mc)$ is the gyrofrequency; $\avrB(\mathbf{x})$ is the background magnetic field; and $c$ is the speed of light. In fusion applications of the electron waves governed by \Eq{eq:Hham}, one typically has $\omega \sim \omega_{\rm p} \sim \Omega$, so we attribute this range, loosely, as the electron-cyclotron range. The eigenvectors of $ \mathsf{D}_{\mathrm{H}}$ are the same as those of $H$, so $H\eta_m = (\Lambda_m + \omega)\eta_m$.

Typically, Berry curvature arises from degeneracy points in the dispersion relation that behave as magnetic monopoles \cite{berry1984quantal}. In the cold plasma model considered here, there are four degeneracy points known as Weyl points \cite{gao2016photonic,fu2021topological}. The spatial locations where $\omega_{\rm p}$ and $\Omega$ are zero also serve as degeneracy points. In addition, Berry curvature has a contribution from infinite $k$ \cite{gao2016photonic}, because the wavevector space is not compact in cold plasmas. Because $\epsilon\ll1$, the system changes slowly on the scale of the wavelength. Thus, we can treat the evolution of a single-mode wave packet as an adiabatic process when mode conversion is absent.

\subsection{SHE in a plasma slab \label{sec:slab}}

First, let us consider waves propagating perpendicular to the magnetic field in a plasma slab, with coordinates denoted $\vec{x} \equiv (x, y, z)$. There are three modes in this case: an O~mode with $\omega > \omega_{\rm p}$ and two X~modes, with $\omega < \omega_{\mathrm{uh}}$ and $\omega > \omega_{\mathrm{uh}}$, respectively, where $\omega_{\mathrm{uh}}$ is the upper-hybrid frequency, $\omega_{\mathrm{uh}} = (\Omega^{2} + \omega_{\mathrm{p}}^{2})^{1/2}$. We assume a homogeneous magnetic field along the $z$~axis, with magnitude $|\avrB| = 0.5\,\text{T}$, so $\Omega \approx 8.8\times 10^{10}\,\text{s}^{-1}$. We also adopt $\omega_{\mathrm{p}}(\mathbf{x})=\omega_{\mathrm{p},0}(1+x/L_{0})$, with $L_{0}=1\,\text{m}$ and $n(x = 0) = 10^{19}\,\text{m}^{-3}$, so $\omega_{\mathrm{p},0} \approx 1.8\times 10^{11}\,\text{s}^{-1}$. The centers of the wave beams simulated pass through $\mathbf{x}_{0}=(0, 0, 0)$, where the wavevector is $\mathbf{k}_{0}=(-200, 0, 0)\,\text{m}^{-1}$, so $\epsilon \sim 0.03$. The ray-tracing simulations are performed, separately for O~waves and X~waves, using \Eqs{eq:rays} and~\eq{eq:UF}. Deviations of the ray trajectory from the $x$~axis in this geometry, if any, are entirely due to the SHE. 

We also compare our XGO ray tracing with FW simulations, which we perform using the finite-difference time-domain method described in Ref.~\cite{fu2022dispersion}. For simplicity, we assume that the system is uniform along the $z$~axis ($\pd_z=0$), so \mbox{2-D} modeling is enough. For the FW simulations, the initial field is taken to be
\begin{gather}
 \Psi = \eta\,\ee^{\ii\mathbf{k}_{0}\cdot\mathbf{x}}\exp\left[-(\mathbf{x}-\mathbf{x}_{0})^{2}/(2\sigma^{2})\right] + \mc{O}(\epsilon),
 \label{eq:0th_initial_wave_packet}
\end{gather}
with $\sigma=3.5\,\text{cm}$. The (nonnegligible) term $\mc{O}(\epsilon)$ is specified in Appendix~\ref{sec:1st_order_initial_condition}. For comparison with the ray-tracing simulations, ``the'' wave coordinate $\vec{x}$ is defined as that of the maximum of the beam action density $\mc{I} \doteq |\Psi|^2$, and the same applies to ``the'' wavevector $\vec{k}$. The spatial grid sizes are chosen to be $\Delta x = \Delta y = 8.0 \times 10^{-5}\,\text{m } \sim 10^{-2} \lambda$ so that the FW simulation can capture the SHE accurately. The temporal grid size is $\Delta t = 7.8 \times 10^{-6}\,\text{ns}$. Then, the phase space velocity of the wave is about $0.1(\Delta x/\Delta t) < \Delta x/\Delta t$, which ensures numerical stability \cite{inan2011numerical}. For a typical case where the XGO equations can be calculated on a laptop within a few minutes, the 2-D FW simulation has to be carried out on a cluster and consumes $\sim 10^{4}$ CPU hours (and 3-D simulations would have been prohibitively expensive.)

\begin{figure}
\centering 
\includegraphics[width=1\columnwidth]{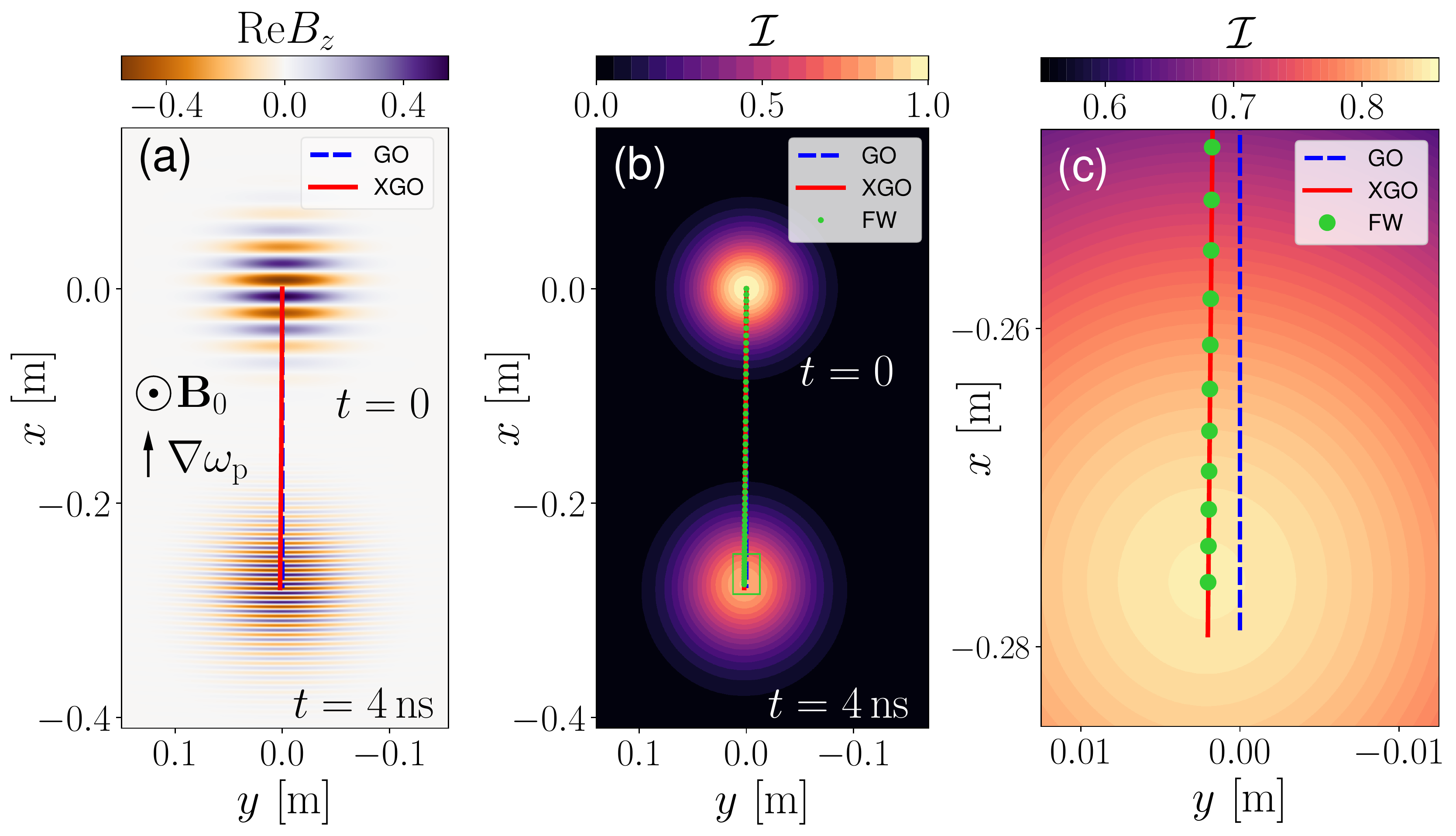} 
\caption{Simulations of an X~wave with $\omega < \omega_{\mathrm{uh}}$ in the $(x, y)$ plane. Shown are snapshots at $t = 0$ and $t = 4\,\text{ns}$. (a)~$\re B_z$, (b)~$\mc{I}$; (c)~a zoom-in on the green rectangle from~(b). The ray trajectories from the GO (\note{dashed} blue curve), XGO (\note{solid} red curve), and FW (green discs) simulations.}
\label{fig:X_wave_combined}
\end{figure}

\begin{figure}[t]
\centering 
\includegraphics[width=7cm]{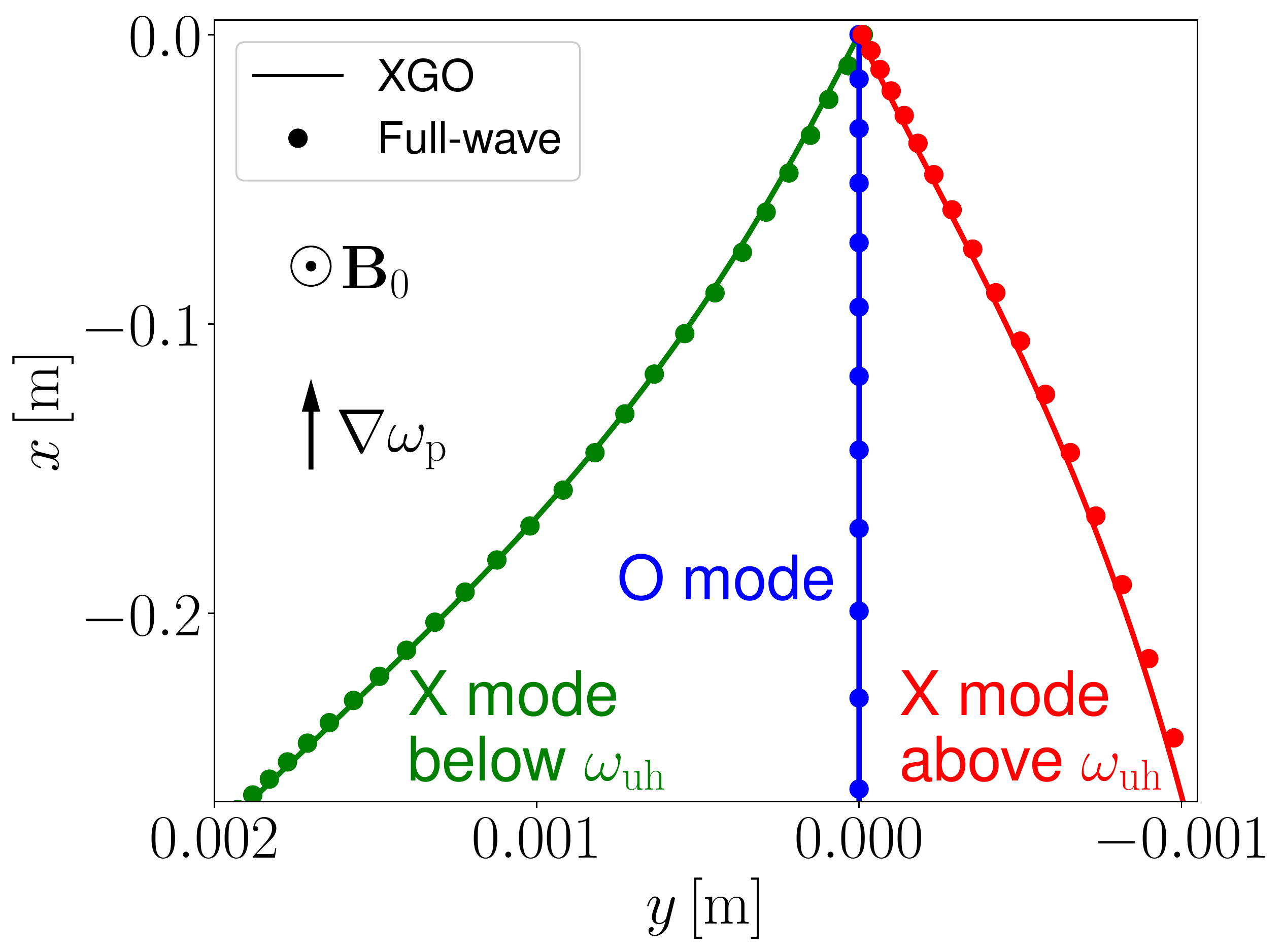} 
\caption{The comparison between XGO (solid curves) and FW (discs) simulation results for the O~wave (blue \note{in the middle}), X~wave with $\omega < \omega_{\mathrm{uh}}$ (green \note{on the left}), and X~wave with $\omega > \omega_{\mathrm{uh}}$ (red \note{on the right}). In GO, all three rays would travel along $y = 0$.}
\label{fig:3_waves}
\end{figure}

As an example, we present typical results of numerical simulations for an X~wave with $\omega < \omega_{\mathrm{uh}}$ in \Fig{fig:X_wave_combined}. The wave packet moves roughly along the $-x$ axis with increasing wavenumber $k_{x}$. The trajectories from GO and XGO are very close to each other yet distinguishable at high resolution. As seen in the zoomed-in plot [\Fig{fig:X_wave_combined}(c)], the separation~$\varsigma$ between them is about $3\,\text{mm}$, and FW simulations are in better agreement with XGO than they are with GO. As could be expected from \Eqs{eq:rays}, this separation constitutes about~1\% of the ray path~$\ell$, so $\varsigma/\ell \sim \epsilon$. For all three modes, the comparison between XGO and FW simulations is presented in \Fig{fig:3_waves}. In GO, all three rays would travel along $y = 0$, so the horizontal displacement of the X-wave rays is entirely due to the SHE. The O~wave does not exhibit the SHE because its polarization vector $\eta$ remains parallel to the $z$-axis and has a fixed phase, so $\pd_\mu \eta \equiv 0$ and $\im [(\pd^\mu \eta^\dag)\pd^\nu \eta] \equiv 0$, so $U_0 \equiv 0$ by \Eq{eq:U0} and $\msf{F} \equiv 0$ by \Eqs{eq:Fs}. Notably, this difference between the SHE for X~and O~waves is consistent with the fact that these modes have different Chern numbers $C$, which are integrals of the Berry curvatures and represent waves' topological invariants \cite{parker2020topological, fu2021topological, qin2022topological}. The O~wave has $C = 0$, the X~wave with $\omega < \omega_{\mathrm{uh}}$ has $C = 2$, and the X~wave with $\omega > \omega_{\mathrm{uh}}$ has $C = -1$.  

\subsection{SHE in a toroidal plasma \label{sec:toroidal}}

The SHE, which accumulates over time, can result in more significant $\varsigma/\ell$ when the group velocity is small. To illustrate this, let us consider the propagation of electromagnetic waves in a toroidal plasma. We assume
\begin{gather}
\omega_{\mathrm{p}}(\mathbf{x}) = \omega_{\mathrm{p,0}}\left[\exp\left({-\frac{(R-R_0)^{2}}{2\sigma_{R}^{2}}-\frac{z^{2}}{2\sigma_{z}^{2}}}\right) + d\right],
\end{gather}
where $R \doteq (x^{2}+y^{2})^{1/2}$, $\sigma_{R}^{2}=0.1\,\text{m}^2$, $\sigma_{z}^{2} = 1\,\text{m}^2$, $d = 0.01$, and the density maximum is located at $(R_0, z_0) = (1, 0)\,\text{m}$, where $\omega_{\mathrm{p}} = \omega_{\mathrm{p,0}} = 1.8 \times 10^{11}\,\text{s}^{-1}$, same as before. We also assume that the magnetic field is toroidal, specifically, aligned with the unit-vector field $(-y/R, x/R, 0)$ and $\Omega(\mathbf{x}) = {\Omega_{0}} R_0/R(\mathbf{x})$, with $\Omega_{0}=8.8\times 10^{10}\,\text{s}^{-1}$, also same as before.

\begin{figure}
\centering \includegraphics[width=8.5cm]{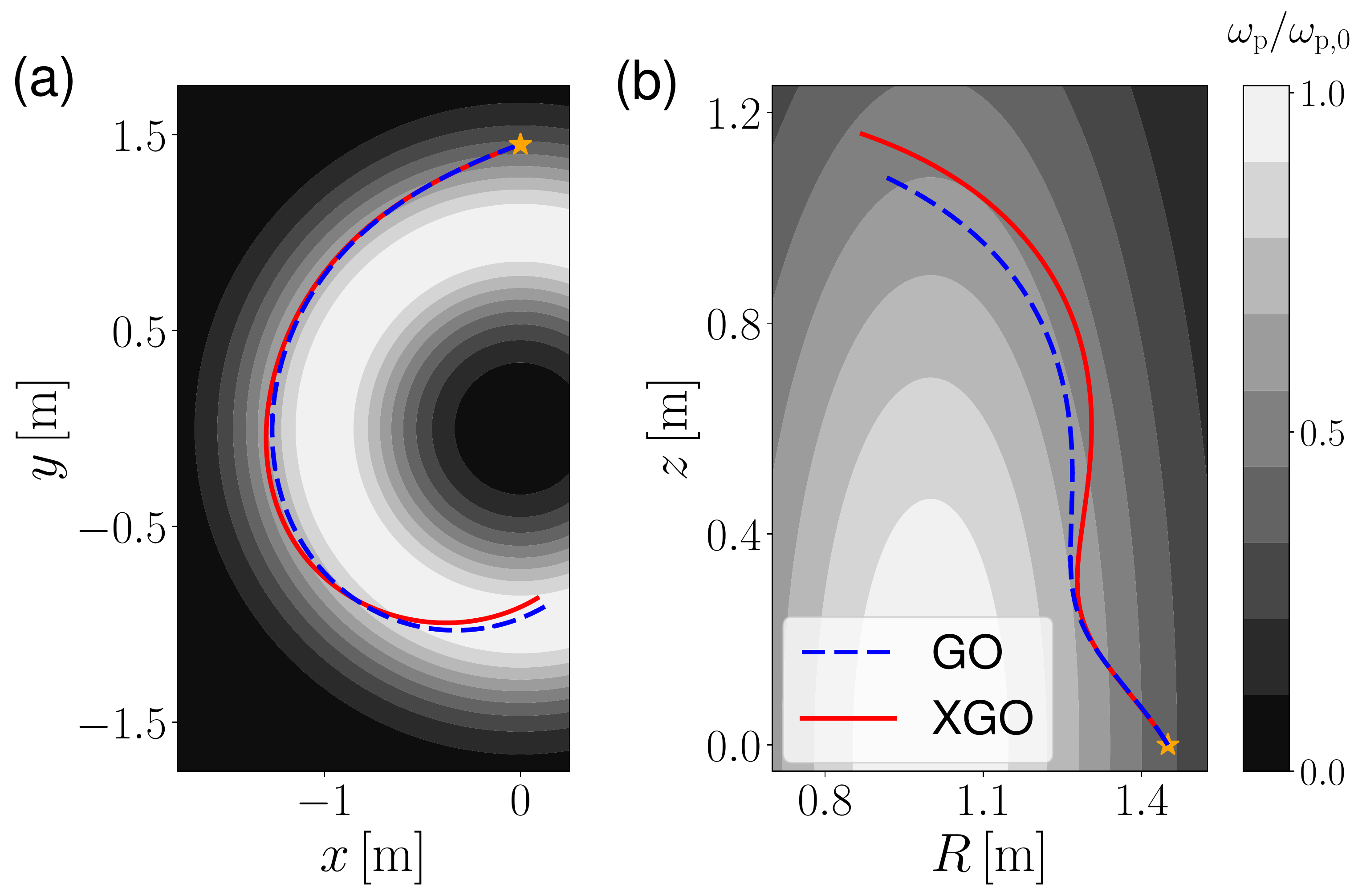} \caption{The propagation of an RF wave in the electron-cyclotron range in (a)~the $(x, y)$ plane and (b)~the $(R, z)$ plane. The gray scale indicates the local density or $\omega_{\rm p}$ (a)~in the plane $z=0$ and (b)~in any toroidal cross section. The stars mark the initial position of the rays. The GO and XGO rays are shown as \note{dashed} blue and \note{solid} red curves, respectively.}
\label{fig:3D_xyzr}
\end{figure}

\begin{figure}
\centering 
\includegraphics[width=8.5cm]{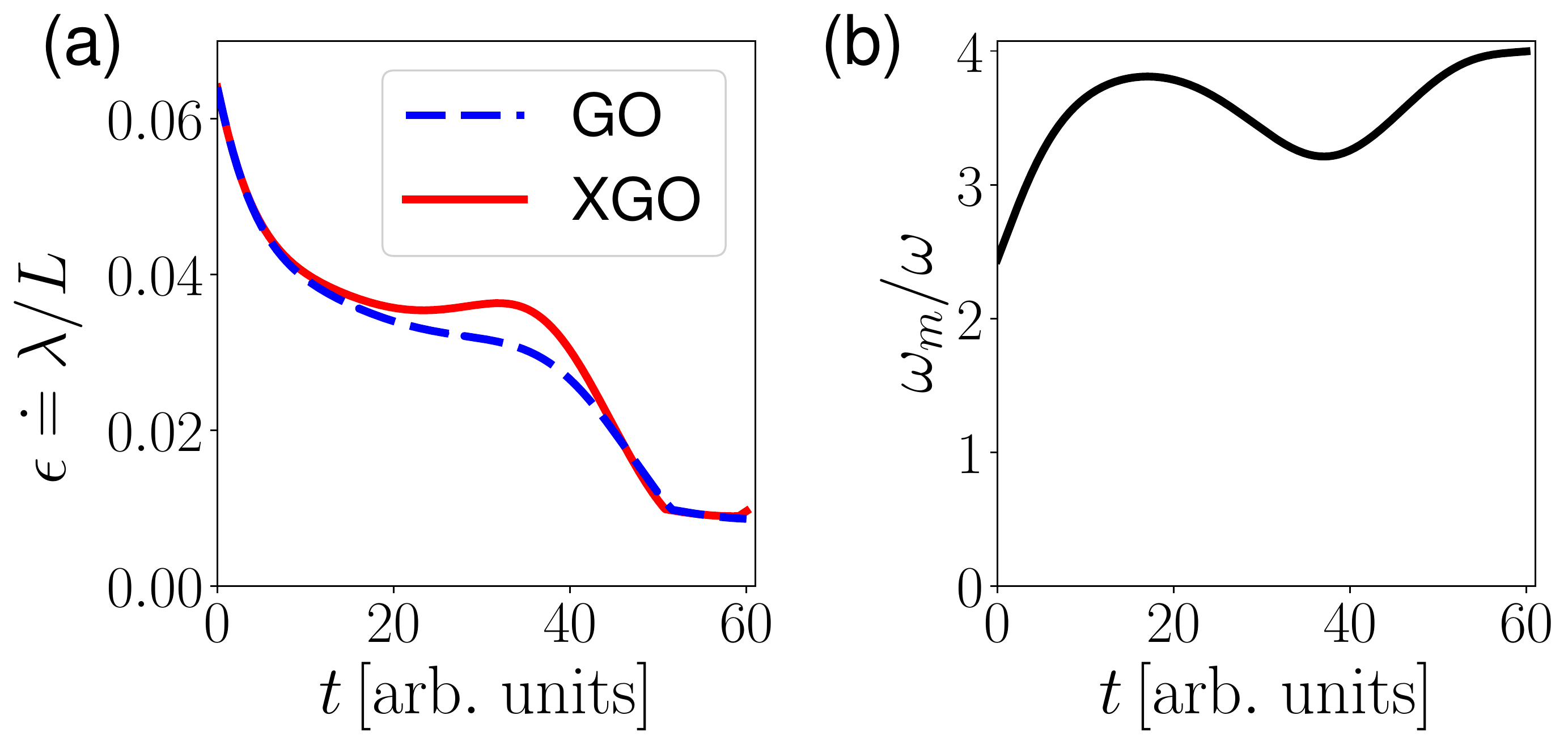} 
\caption{(a)~The local GO parameter $\epsilon$ along the GO and XGO trajectories. The local length scale is calculated as $L(\mathbf{x}) \doteq \min (|\omega_{\mathrm{p}}/\nabla\omega_{\mathrm{p}}|,|\boldsymbol{\Omega}/\nabla\boldsymbol{\Omega}|)$. \note{The GO and XGO results are shown as dashed blue and solid red curves, respectively.} (b)~The frequency $\omega_m$ of the passive mode closest to the active mode evaluated on the XGO trajectory $(\vec{x}(t), \vec{k}(t))$, in units $\omega$, as a function of the ray path. Since $\omega_m/\omega - 1 \sim 1$, mode conversion is not to be expected.}
\label{fig:3D_lambda_omega}
\end{figure}

Consider a wave starting at $\mathbf{x}_{0}=(0,1.45,0)\,\text{m}$ with initial wavevector $\mathbf{k}_{0}=(-330,150,-250)\,\text{m}^{-1}$ and $\omega \approx 3.1 \times 10^{10}\,\text{s}^{-1}$, which is the lowest-frequency mode in the system. The numerical results are presented in \Fig{fig:3D_xyzr} and show that the separation between the XGO and GO trajectories is as large as $\varsigma \sim 0.1\,\text{m}$, which is about ten wavelengths. FW simulations for this case would have to be 3-D, \ie computationally expensive, so they are not reported. Instead, we have calculated the parameters that determine XGO applicability. We have found that $\epsilon \lesssim 0.06 \ll 1$ [\Fig{fig:3D_lambda_omega}(a)] and the passive mode closest to the active mode remains nonresonant on the whole ray trajectory [\Fig{fig:3D_lambda_omega}(b)]. This means that breaking of the XGO ordering and mode conversion that we have ignored are indeed not to be expected. In other words, our simulations are well within the XGO validity regime, hence the predicted large value of $\varsigma$ is not an artifact. Such a large deviation can significantly affect resonant absorption of radiofrequency waves in fusion applications \cite{prater2008benchmarking}. Therefore, typical ray-tracing codes that are based on GO instead of XGO  are at risk of missing important physics, even if \textit{usually} the SHE is less pronounces than in our example.

\section{Conclusions \label{sec:conclusion}}

Here, we present the first systematic study of the SHE for magnetized plasmas. We start with the XGO formulation of the SHE and derive a gauge-invariant form of the ray equations that describe SHE for general waves [\Eq{eq:rays}]. We also express the right-hand side in a form that is better suited for simulations due to its ``numerical gauge invariance'' [\Eqs{eq:UF}]. Then, we perform ray-tracing simulations based on these equations for electromagnetic waves in a cold magnetized collisionless electron plasma slab and compare them with 2-D FW simulations. We show that the FW simulations are in better agreement with XGO, which retains the SHE, than they are with GO simulations, which ignores the SHE. Finally, we present an example of a large SHE in toroidal plasma, where a wave beam deviates from the GO trajectory by a distance of roughly ten wavelengths. In fusion devices, such a large deviation can significantly affect resonant absorption of radiofrequency waves. Therefore, typical ray-tracing codes that are based on GO instead of XGO  are at risk of missing important physics, even if \textit{usually} the SHE is less pronounces than in our example. 

\begin{acknowledgments}
   This research was supported by the Department of Energy through contract No.\ DE-AC02-09CH11466.  Y.\ F.\ thanks Peifeng Fan for fruitful discussions. 
\end{acknowledgments}

\appendix

\section{Gauge invariance of the ray Lagrangian \label{sec:gauge}}

In this section, we prove the gauge invariance of the ray Lagrangian defined in Eq.~(\ref{eq:L}):
\begin{gather}
    \mc{L} = \left[\dfrac{k_\mu}{2} - \mathsf{A}_{\mu}^{(x)}\right]\dot{x}^\mu - \left[\dfrac{x^\mu}{2} + \mathsf{A}^{\mu}_{(k)}\right]\dot{k}_\mu - (\Lambda - U_0),
    \notag
\end{gather}
where
\begin{gather}
U = U_0 - \mathsf{A}_{\mu}^{(x)}\,\pd^\mu\Lambda + \mathsf{A}_{(k)}^{\mu}\,\pd_\mu \Lambda,\\
U_0 \doteq \im \left[ \left(\pd^\mu\eta^{\dag}\right) \mathsf{D}_{\mathrm{H}} \left(\pd_\mu\eta\right) \right],
\label{eq:U0_a}\\
\mathsf{A}_{\mu}^{(x)} \doteq \im \left(\eta^{\dag} \pd_\mu\eta \right),
\quad
\mathsf{A}_{(k)}^{\mu} \doteq \im\left(\eta^{\dag} \pd^\mu\eta \right).
\label{eq:As_a}
\end{gather}
As a reminder, we assume the notation $\pd_\mu \doteq \pd/\pd x^{\mu}$ and $\pd^\mu \doteq \pd/\pd k_{\mu}$, and $\eta$ denotes a unit polarization vector of the active eigenmode, which is also an eigenvector of $\mathsf{D}_\mathrm{H}$:
\begin{gather}
    \mathsf{D}_\mathrm{H}\eta = \Lambda \eta, 
    \quad
    \eta^\dag \eta = 1,
    \quad
    \Lambda = \eta^\dag \mathsf{D}_\mathrm{H}\eta,
\end{gather}
and $\Lambda$ is the corresponding eigenvalue.

Let us consider a gauge transformation $\eta\to \ee^{\ii\varphi}\eta$, where $\varphi(x^\mu, k_\mu)$ is a real function. This transformation does not affect $\Lambda$, while $U_0$ is transformed as follows:
\begin{gather}
U_0 \to \im \big[ 
        (\pd^\mu \eta^\dag) \mathsf{D}_\mathrm{H} (\pd_\mu \eta) 
        + (\pd^\mu \varphi) (\pd_\mu \varphi) \Lambda + \Phi \big],
                \label{eq:U0_1_a} 
\\
\Phi = \ii \Lambda \big[ (\pd_\mu \varphi) (\pd^\mu \eta^\dag) \eta
        - (\pd^\mu \varphi) \eta^\dag (\pd_\mu \eta) \big]. 
\label{eq:U0_2_a}
\end{gather}
Since both $\varphi$ and $\Lambda$ are real, the term $(\pd^\mu \varphi) (\pd_\mu \varphi) \Lambda$ is real and therefore does not contribute to $U_0$. Also,
\begin{gather}\label{eq:puim_a}
0 = \pd(\eta^\dag \eta) = (\pd\eta^\dag)\eta + \eta^\dag (\pd \eta),
\end{gather}
so $(\pd \eta^\dag)\eta$ is purely imaginary. Here and further, $\pd$ denotes a derivative with respect to any coordinate in the ray phase space $(\vec{x}, \vec{k})$. Hence, $\Phi$ is real and therefore does not contribute to $U_0$ either. Then, according to \Eq{eq:U0_1_a}, the gauge transformation leaves $U_0$ intact. 

Now let us consider the Berry connections $\mathsf{A}$. These functions transform as follows:
\begin{gather}
\mathsf{A}^{(x)}_\mu
\to \im\left( \eta^\dag \pd_\mu \eta  + \ii \pd_\mu \varphi \right)
= \mathsf{A}^{(x)}_\mu + \pd_\mu \varphi,
\end{gather}
\begin{align}
    \mathsf{A}_{\mu}^{(x)} \dot{x}^{\mu} & + \mathsf{A}_{(k)}^{\mu} \dot{k}_{\mu} \notag\\
    & \to \mathsf{A}_{\mu}^{(x)} \dot{x}^{\mu} + \mathsf{A}_{(k)}^{\mu} \dot{k}_{\mu} 
    +  (\pd_\mu \varphi) \dot{x}^{\mu} +  (\pd^\mu \varphi) \dot{k}_{\mu} \notag \\
    & = \mathsf{A}_{\mu}^{(x)} \dot{x}^{\mu} + \mathsf{A}_{(k)}^{\mu} \dot{k}_{\mu} +  \dot{\varphi}.
\end{align}
The extra term $\dot\varphi$ is a total time derivative and thus does not contribute to the equation of motion. Therefore, up to this insignificant full time derivative, the non-canonical Lagrangian \eq{eq:L} is gauge-invariant and so are the corresponding ray equations.

\section{Derivation of \texorpdfstring{$\boldsymbol{\mathsf{F}}$}{F} and \texorpdfstring{$\boldsymbol{U_0}$}{U0}
\label{sec:gauge_invariant_formula}}

The formulas for $\mathsf{F}$ and $U_0$ are well-known in condensed matter physics, but not so well known for general waves and in plasma physics in particular, so it is instructive to present their general derivation. Here, we do so by adapting the corresponding calculation from Ref.~\cite{berry1984quantal}.

Because $\mathsf{D}_\mathrm{H}$ is a Hermitian matrix, one can choose its orthonormal eigenvectors to form a complete basis. Let us consider any one of them, $\eta_n$, and decompose its derivative (with respect to any given parameter) in this basis:
\begin{align}
\pd \eta_n =  (\eta_n^\dag \pd\eta_n) \eta_n + \sum_{s \ne n} (\eta_s^\dag \pd\eta_n) \eta_s.
\label{eq:deta_a}
\end{align}
To calculate the coefficients in the latter sum, let us differentiate $\mathsf{D}_\mathrm{H}\eta_n = \Lambda_n \eta_n$ and multiply the result by $\eta_m^\dag$ with $m\neq n$ from the left. This gives
\begin{gather}
\eta_m^\dag \pd \eta_n = \dfrac{\eta_m^\dag (\pd \mathsf{D}_\mathrm{H} ) \eta_n}{\Lambda_n - \Lambda_m} 
\quad
(m\neq n),
\end{gather}
assuming $\Lambda_n \neq \Lambda_m$. Hence, one can rewrite \Eq{eq:deta_a} as
\begin{gather}
\pd \eta_n = \left(\eta_n^\dag \pd\eta_n\right) \eta_n + 
\sum_{m\neq n} \dfrac{\eta_m^\dag (\pd \mathsf{D}_\mathrm{H} ) \eta_n }{\Lambda_n - \Lambda_m} \eta_m.
\label{eq:eigen_derivative_a}
\end{gather}
Plugging this into \Eq{eq:U0_a} yields
\begin{gather}
U_0 = \im
\sum_{m\neq n} \dfrac{\eta_n^\dag (\pd^\mu \mathsf{D}_\mathrm{H}) \eta_m \eta_m^\dag (\pd_\mu \mathsf{D}_\mathrm{H}) \eta_n}{(\Lambda_m - \Lambda_n)^2/\Lambda_m}.
\label{eq:U02}
\end{gather}
Similarly, plugging \Eq{eq:eigen_derivative_a} into
\begin{gather}
\mathsf{F}^{\mu\nu}\doteq 2\im[(\partial^\mu\eta_n^{\dag}) \partial^\nu\eta_n]
\end{gather}
and making use of the fact that $(\pd \eta_n^\dag)\eta_n$ is purely imaginary [cf.\ \Eq{eq:puim_a}], leads to
\begin{gather}\label{eq:FF_a}
\mathsf{F}^{\mu\nu} = 2\im \sum_{m\neq n} 
\dfrac{\eta_n^\dag (\pd^\mu \mathsf{D}_\mathrm{H}) \eta_m \eta_m^\dag (\pd^\nu \mathsf{D}_\mathrm{H}) \eta_n}{(\Lambda_n-\Lambda_m)^2}.
\end{gather}

Finally, let us choose $\eta_n$ to be the polarization vector of the active mode, $\eta_n = \eta$, so $\Lambda_n = \Lambda$. By the dispersion relation, one has $\Lambda = \mc{O}(\epsilon)$, so within the assumed accuracy \Eqs{eq:U02} and \eq{eq:FF_a} can as well be written as follows:
\begin{gather}
U_0 = \im
\sum_{m\neq n} \dfrac{\eta_n^\dag (\pd^\mu \mathsf{D}_\mathrm{H}) \eta_m \eta_m^\dag (\pd_\mu \mathsf{D}_\mathrm{H}) \eta_n}{\Lambda_m},
\\
\mathsf{F}^{\mu\nu} = 2\im
\sum_{m\neq n} \dfrac{\eta_n^\dag (\pd^\mu \mathsf{D}_\mathrm{H}) \eta_m \eta_m^\dag (\pd^\nu \mathsf{D}_\mathrm{H}) \eta_n}{\Lambda_m^2}.
\end{gather}
Similar formulas for ${\mathsf{F}^\mu}_\nu$, ${\mathsf{F}_\mu}^\nu$, and $\mathsf{F}_{\mu\nu}$ are derived in the same way.

\section{Construction of wave packets \label{sec:1st_order_initial_condition}}

Within XGO, terms $\mc{O}(\epsilon)$ in expressions for the field are not negligible \cite{dodin2019quasioptical}. In particular, they must be retained when constructing initial wave packets for full-wave simulations if the spin Hall effect must be retained. In this section, we derive an explicit approximation for these terms and apply it to cold-plasma waves.

A slow complex envelope $\psi(x^\mu)$ is governed by
\begin{equation}
    \hat{\mc{D}}\psi=0, \label{eq:envelop_eq_a}
\end{equation}
where $\hat{\mc{D}}$ is approximated by Eq.~(1) of the main article up to insignificant terms $\mc{O}(\epsilon^2)$. This operator can be represented as $\hat{\mc{D}}\approx \msf{D}_\mathrm{H}(x^{\mu}, \bar{k}_{\mu}(x^\mu)) + \hat{\mathscr{D}}$, where $\msf{D}_\mathrm{H} = \mc{O}(1)$ and $\hat{\mathscr{D}} = \mc{O}(\epsilon)$ under the assumptions specified in the main article.  Then, one can expand $\psi$ in the basis of orthonromal eigenvectors of $\mathsf{D}_{\rm H}(x^{\mu}, \bar{k}_{\mu}(x^\mu))$ as
\begin{equation}
\psi(x^\mu) = \sum_m \eta_m\left(x^{\mu}, \bar{k}_{\mu}(x^\mu)\right)a_m(x^\mu),
\end{equation}
where $a_n$ are scalar amplitudes. Then, Eq.~(\ref{eq:envelop_eq_a}) gives
\begin{equation}\label{eq:la_a}
\sum_m  \Lambda_m \eta_m a_m = \mathcal{O}(\epsilon),
\end{equation}
whence $\Lambda_m a_m = \mathcal{O}(\epsilon)$ for each $m$. Consider the region where $\msf{D}_\mathrm{H}$ has only one eigenvalue that is close to zero, say $\Lambda_n = \mathcal{O}(\epsilon)$, \ie only the $n$th mode is on-shell (propagating wave that satisfies a dispersion relation). Such mode may have $a_n = \mathcal{O}(1)$ and is called an active mode. The remaining modes, for which $\Lambda_m = \mathcal{O}(1)$, have to have $a_m = \mathcal{O}(\epsilon)$ and are called passive modes \cite{dodin2019quasioptical}. 

Although small, passive modes must be accounted in simulations that are intended to capture the SHE. In particular, the initial wave envelope $\psi(t = 0)$ must be constructed with accuracy not less than $\mathcal{O}(\epsilon)$. To do, let us represent the envelope as $\psi = \eta_n a_n + \psi_\perp$,
\begin{align}
\psi_\perp \doteq \sum_{m \neq n} \eta_m  a_m = \mathcal{O}(\epsilon). 
\end{align}
Then,
\begin{align}
(\msf{D}_\mathrm{H} + \hat{\mathscr{D}}) (\eta_n a_n + \psi_\perp ) = 0,
\end{align}
where an error $\mathcal{O}(\epsilon^2)$ in the approximation of $\hat{\mc{D}}$ has been ignored. Since $\hat{\mathscr{D}}\psi_\perp = \mathcal{O}(\epsilon^2)$, this term can be ignored as well. Then, multiplying the equation by $\eta_m^\dag$ with $m\neq n$ from the left, one obtains
\begin{equation}
a_m = - \dfrac{\eta_m^\dagger \hat{\mathscr{D}} (\eta_n a_n) }{ \Lambda_m} + \mathcal{O}(\epsilon^2).
\end{equation}
Therefore, up to $\mathcal{O}(\epsilon)$, the total wave field $\Psi$ can be expressed in the following form parameterized by a single function, $a_n$:
\begin{equation}
\Psi(\mathbf{x}) = \mathrm{e}^{\mathrm{i}\theta} \left( 
    a_n \eta_n - \sum_{m\neq n} \dfrac{\eta_m^\dagger \hat{\mathscr{D}} (\eta_n a_n) }{ \Lambda_m}
\right).
\end{equation} 

For cold plasma, the dispersion operator is $\hat{D} = \hat{H} - \mathrm{i}\partial_t$, where $\hat{H}(\mathbf{x}, -\ii\partial_{\mathbf{x}})$ is defined in Eq.~(12) of the main article. Using $\theta=k_\mu x^\mu$ and $\msf{D}_\mathrm{H} = H(\mathbf{x},\mathbf{k}) - \omega$, the operator $\hat{\mathcal{D}}=\msf{D}_\mathrm{H}+\hat{\mathscr{D}}$ can be calculated exactly:
\begin{equation}
    \hat{\mathscr{D}} = 
\begin{pmatrix}
    0 & 0 & 0 \\
    0 & 0 & \ii c \, \pd_\mathbf{x}\times \\
    0 & -\ii c \, \pd_\mathbf{x}\times & 0
\end{pmatrix} - \ii\pd_t,
\end{equation} 
where $\pd_t = 0$ for stationary waves.

\bibliographystyle{apsrev2}
\bibliography{spin_hall.bbl}

\begin{thebibliography}{43}
\bibitem [{foo()}]{foot:quasi}%
  \BibitemOpen
  \href@noop {} {}\bibinfo {note} {Quasioptical schemes
  \cite{dodin2019quasioptical, ref:balakin07a} do not infer the envelope
  evolution from the rays but rather evolve the envelope independently. As a
  result, the envelope is allowed to drift away from the reference ray and SHE
  is automatically accounted for as long as this effect is not too
  large.}\BibitemShut {Stop}%
\end{thebibliography}
  
\end{document}